\documentclass[aps,twocolumn,superscriptaddress,floatfix,preprintnumbers,nofootinbib]{revtex4-1}

\usepackage{graphicx}
\usepackage{dcolumn}
\usepackage{bm}

\usepackage[utf8]{inputenc}
\usepackage{ulem}
\usepackage[margin=1in]{geometry}
\usepackage{tikz}
\usepackage[compat=1.1.0]{tikz-feynman}
\usepackage[T1]{fontenc}
\usepackage[caption=false]{subfig}
\usepackage{amssymb}
\usepackage{xcolor}
\usepackage{amsmath}
\usepackage{comment}
\usepackage{natbib}
\usepackage[pdftex,bookmarks,linktocpage,pdfpagelabels,plainpages=false,hyperfigures,linkcolor=blue,citecolor=blue]{hyperref}

\hypersetup{colorlinks=true}

\begin{document}

\preprint{MI-HET-779}

\title{Axion-Like Particle Production at Beam Dump Experiments with Distinct Nuclear Excitation Lines}

\author{Loyd Waites}
\affiliation{%
 Physics Department, Massachusetts Institute of Technology
}%
\email{lwaites@mit.edu}

\author{Adrian Thompson}
\affiliation{Mitchell Institute for Fundamental Physics and Astronomy, Department of Physics and Astronomy, Texas A\&M University, College Station, TX 77845, USA}

\author{Adriana Bungau}%
\affiliation{%
 Physics Department, Massachusetts Institute of Technology
}%
\author{Janet M. Conrad}
\affiliation{%
 Physics Department, Massachusetts Institute of Technology
}%

\author{Bhaskar Dutta}
\affiliation{Mitchell Institute for Fundamental Physics and Astronomy, Department of Physics and Astronomy, Texas A\&M University, College Station, TX 77845, USA}

\author{Wei-Chih Huang}
\affiliation{Mitchell Institute for Fundamental Physics and Astronomy, Department of Physics and Astronomy, Texas A\&M University, College Station, TX 77845, USA}

\author{Doojin Kim}
\affiliation{Mitchell Institute for Fundamental Physics and Astronomy, Department of Physics and Astronomy, Texas A\&M University, College Station, TX 77845, USA}

\author{Michael Shaevitz}
 
\affiliation{
 Physics Department, Columbia University
}%

\author{Joshua Spitz}
\affiliation{%
 Physics Department, University of Michigan
}%


\date{\today}

\begin{abstract}
Searches for axion-like particles (ALPs) are motivated by the strong CP problem in particle physics and by unexplained dark matter in astrophysics. In this letter, we discuss novel ALP searches using monoenergetic nuclear de-excitation photons from a beam dump, using IsoDAR as an example. We show that IsoDAR can set limits that close a gap in traditional QCD axion searches using the ALP-photon coupling, as well as provide sensitivity to large regions of new parameter space in models where ALPs couple to nucleons and electrons. We also show how isotope decay-at-rest experiments may be designed to improve potential ALP production and optimize detection sensitivity.
\end{abstract}

\maketitle

\noindent \textbf{Introduction}.
 Axions were originally motivated during the breaking of Peccei-Quinn symmetry, which had been proposed to solve the  strong CP problem~\cite{Peccei:1977hh,Wilczek:1977pj,Weinberg:1977ma,Preskill:1982cy,Abbott:1982af,Dine:1982ah,Duffy:2009ig,Marsh:2015xka,Battaglieri:2017aum}, the parameter space for such pseudoscalar axion-like particles (ALPs) has since been expanded to include the pseudo-Nambu-Goldstone Bosons (pNGBs) of other broken symmetries like Majorons and Familons~\cite{Irastorza:2018dyq}, ALP dark matter~\cite{Chadha-Day:2021szb}, heavy ALPs in the MeV to GeV mass scales~\cite{Hook:2019qoh,Valenti:2022tsc,Kelly:2020dda}, and so called ``axiverse'' pseudoscalars which may number in the tens to hundreds from string theory compactification scenarios (see, e.g., refs.~\cite{Arvanitaki:2009fg,Cicoli:2012sz}). Designing an experiment for such a broad parameter space requires high-intensity sources of ALP flux and strong background rejection. 


ALPs have been searched for using a variety of methods both in laboratory and using astrophysical observations~\cite{Asztalos:2001tf,Du:2018uak, Kahn:2016aff,Salemi:2019xgl,Brubaker:2016ktl,Droster:2019fur,JacksonKimball:2017elr,Zioutas:1998cc,Anastassopoulos:2017ftl,Irastorza:2013dav,IAXO:2019mpb,Melissinos:2008vn,DeRocco:2018jwe,Obata:2018vvr,Liu:2018icu,Volpe:2019nzt,Dusaev:2020gxi,Banerjee:2020fue,Feng:2018noy,Berlin:2018bsc,Akesson:2018vlm,Berlin:2018pwi,Alekhin:2015byh,Bonivento:2019sri,Dent:2019ueq,AristizabalSierra:2020rom,Chang:2006ug,Oka:2017rnn,Armengaud:2013rta,Armengaud:2018cuy,PhysRevD.101.052008,Aprile:2020tmw,Dent:2020jhf,Fu_2017,Moriyama:1995bz,Krcmar:1998xn,Krcmar:2001si,Derbin:2009jw,Gavrilyuk:2018jdi,Creswick:2018stb,Li:2015tsa,Li:2015tyq,Benato:2018ijc,Dent:2021jnf,Dent:2020qev, Carenza:2021alz, Galanti:2018nvl,Tavecchio:2012um,Galanti:2015rda,Ayala:2014pea,Fermi-LAT:2016nkz,Conlon:2013txa,Conlon:2015uwa,Conlon:2017qcw,Raffelt:1987im}. Most relevant for the IsoDAR experiment~\cite{alonso2022isodar, alonso2022neutrino,alonso2021isodar,winklehner2018high} discussed here, ALP searches have been proposed at several beam dumps and nuclear reactors using neutrino detectors~\cite{Kelly:2020dda, Berryman:2019dme, ichep}. In GeV-scale beam dump experiments, ALP production from monoenergetic nuclear excitations may be washed out, as the GeV-scale proton beam will leave the majority of its energy footprint in the continuum spectra of hadronic and electromagnetic particle cascades before reaching the characteristic MeV scale of nuclear excitations. However, MeV scale beam dumps offer the capability to utilize such transitions to study ALPs as the dominant production modes via couplings to nucleons, and inducing a sharply peaked signal that may be distinguished from backgrounds. We present the capability of using an MeV-scale proton beam to probe new parameter space of ALPs sourced from electromagnetic secondaries as well as nuclear excited states in the beam target environment. 

\begin{figure}[t!]
    \centering
    \includegraphics[width=0.35\textwidth]{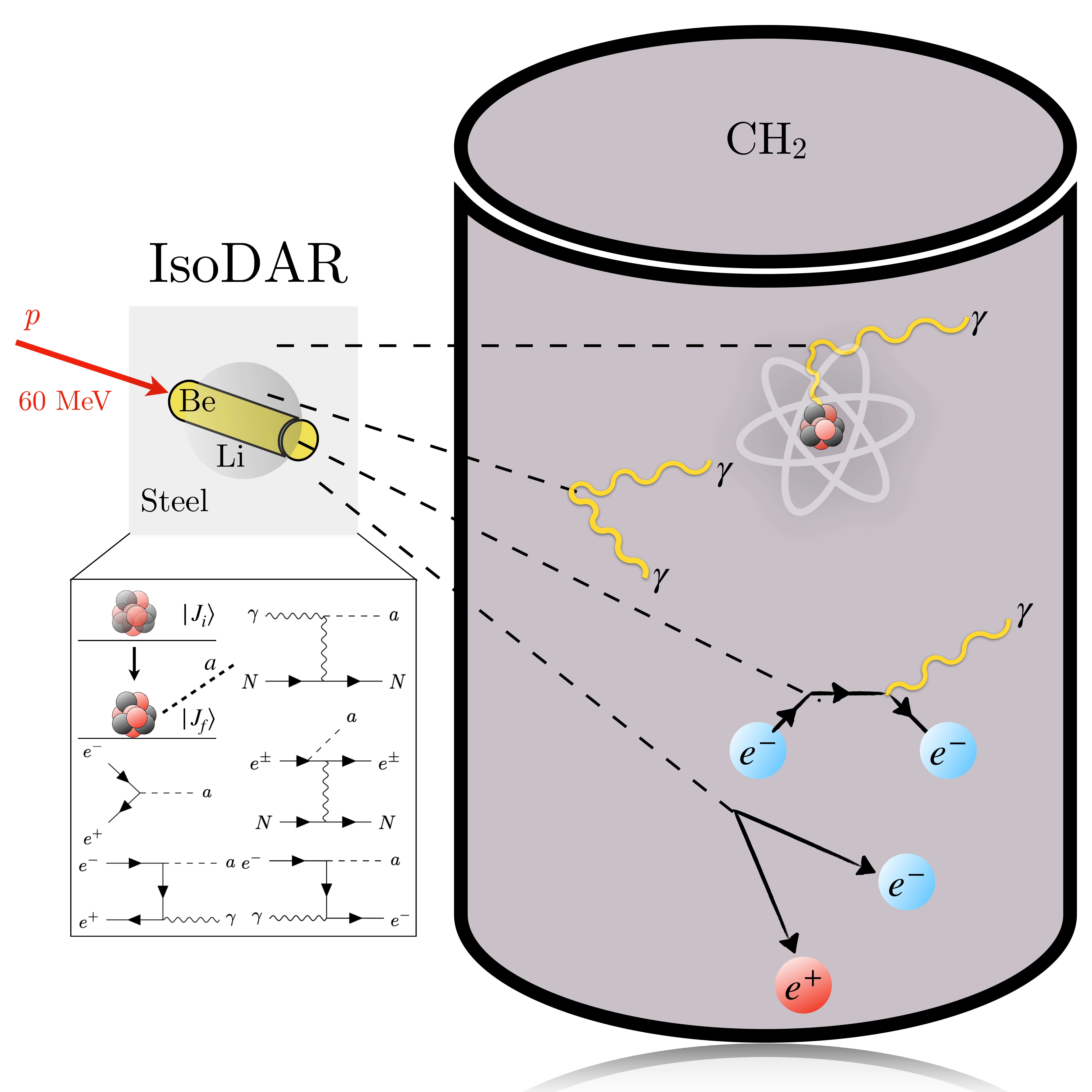}
    \caption{ALP production from the IsoDAR@Yemilab target, and detection via electron, nucleon, and photon couplings leading to $\gamma$, $\gamma\gamma$, $\gamma e^-$, and $e^+ e^-$ final states in the LSC detector.}
    \label{fig:axion}
\end{figure}

The IsoDAR (Isotope Decay-at-Rest) experiment will use a high power cyclotron to produce 10 mA of protons at 60 MeV ($7.9\cdot10^{24}$ protons on target over 5 years, including downtime). The protons impinge on a beryllium target surrounded by a $^7$Li/beryllium sleeve, which through nuclear processes produce electron antineutrinos. The proton-target interactions simultaneously produce photons, mainly from the de-excitation of the nuclei. The entire photon distribution has been modeled in GEANT4 and shown to be isotropic. This is due to the dominance of secondary interactions producing photons. In this paper, we investigate how these photons and nuclear transitions can produce ALPs and be detected with the Liquid Scintillator Counter (LSC; 2.26~kton, 15~m diameter, 15~m height) detector at Yemilab~\cite{seo2019neutrino}, 17~m center-to-center from the target~\cite{alonso2021isodar, alonso2022isodar, Alonso_2022}.

We demonstrate that the IsoDAR at Yemilab experiment can explore new parameter space for ALPs. It can also provide laboratory-based, model-independent constraints over parameter space excluded only by astrophysical considerations.

\noindent \textbf{ALP Production in the IsoDAR Target.}

To understand the ALP production channels we simulate the nuclear interactions inside the IsoDAR target using GEANT4~\cite{Agostinelli:2002hh}. The photon flux is calculated using the \texttt{QGSP\_BIC\_AllHP} physics list, and based on a detailed IsoDAR target and sleeve geometry~\cite{alonso2022isodar, Alonso_2022}. In addition to modeling the primary nuclear interactions from beam collision, GEANT4 accounts for secondary photon production from processes such as nuclear and atomic excitations. This includes secondary neutron scattering and cascade photons which are crucial processes in IsoDAR~\cite{alonso2022isodar,Alonso_2022, alonso2022neutrino, alonso2021isodar}. 

We begin by considering electromagnetically coupled ALPs, described by 
\begin{equation}
    \mathcal{L}_{a\gamma} = -\frac{1}{4}g_{a\gamma} a F_{\mu\nu} \tilde{F}^{\mu\nu}
    \label{eq:LagGamma}
\end{equation}

We estimate the ALP flux by finding the probability of ALP production per photon and sum over the flux and each target layer to find the total ALP flux. The integrated cross section for Primakoff scattering has been reported in refs.~\cite{Tsai:1986tx,Avignone:1988bv,Creswick:1997pg,Avignone:1997th}.

Similarly, electron ALP couplings of the Yukawa form 
\begin{equation}
    \mathcal{L}_{ae} \supset i g_{ae} a \overline{\psi}_e \gamma^5 \psi_e
\end{equation}
may be searched using several channels. In the IsoDAR target, products produced in electromagnetic showers may interact with the target material producing an ALP flux through an ALP-electron coupling. Photons impinging on target electrons at rest can undergo Compton-like scattering ($\gamma e^- \to a e^-$), while electrons and positrons may source resonantly produced ALPs ($e^+ e^- \to a$), or ALPs through associated production ($e^+ e^- \to a \gamma$), and bremsstrahlung ($e^\pm N \to e^\pm N a$). These channels have been studied recently in the context of proton beam targets in refs.~\cite{Capozzi:2021nmp,CCM:2021lhc,Bhattarai:2022mue}.
For ALPs produced from Compton scattering, we convolve the scattering cross section with the photon flux in the IsoDAR target. For ALPs produced from bremsstrahlung, resonance, and associated production, however, the energy loss of the electrons and positrons in the IsoDAR target must also included. For $e^+ e^- \to a$ resonant production from positrons impinging on IsoDAR target electrons, the ALP rate is
\begin{align}
    \Phi_a &\propto \int_{E_+^{min}}^{E_+^{max}} \int^{E_+} \int_0^T\frac{d\Phi_{e^+}}{dE_+} \nonumber \\
    & \times I(t, E_+, E^\prime) \sigma(E^\prime) dt dE^\prime dE_+
\end{align}
where $I(t, E_i, E_f) = \frac{\theta(E_i - E_f)}{E_i \Gamma (4 t/3)} (\ln E_i/E_f)^{4t/3 - 1}$ is the energy loss smearing function for the electron/positron radiation length $t$ and target radiation thickness $T$~\cite{PhysRevD.34.1326}.

Similar fluxes are derived for associated production and bremsstrahlung, except the differential energy cross section is used, resulting ALP flux takes on a continuous energy spectrum.

\begin{figure}[tbh]
    \centering
    \includegraphics[width=.98\columnwidth]{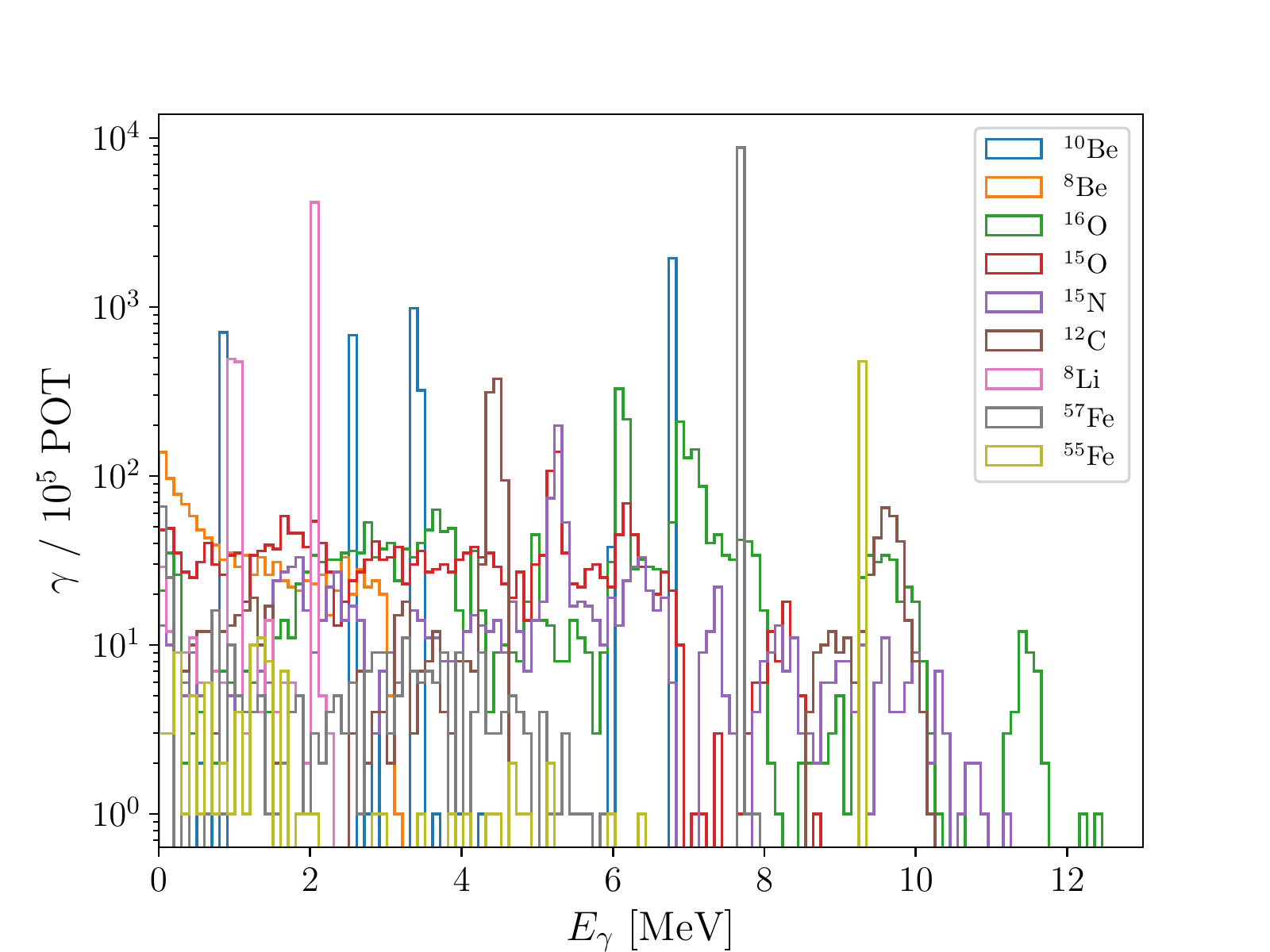}
    \caption{Photon spectrum generated from the GEANT4 simulation of IsoDAR. Photon sources include both continuum and discrete channels, notably neutron capture and inelastic processes involving $d$, $p$, $n$, and $\alpha$ channels.}
    \label{fig:GEANTplot}
\end{figure}

We consider models that predict the decay rate ratio  $\Gamma_a / \Gamma_\gamma$ for nuclear decay $N^* \rightarrow N + a/\gamma$. A pseudoscalar ALP $a$ is associated with MJ transitions (magnetic multipole transitions with angular momentum J). Magnetic multipole transitions, e.g. magnetic dipole (M1) and quadrupole (M2), have angular momentum change $\Delta I= J$ and parity change $\Delta \pi=(-1)^{J+1}$.
The coupling is given by
\begin{equation}
\mathcal{L}_{aN} = ia \bar{\psi}_N\gamma_5({g^0_{aNN}}+g^1_{aNN}\tau_3)\psi_N 
\end{equation}
where $\psi_N= \left( \begin{array}{c}
     p \\
     n 
\end{array} \right)$. The branching ratio for the transitions to ALPs is~\cite{Avignone:1988bv}
\begin{eqnarray}
\label{eq:br}
\left( \frac{\Gamma_a}{\Gamma_\gamma}\right)_{\text{MJ}} &=& \frac{1}{\pi \alpha} \frac{1}{1+\delta^2} \frac{J}{J+1} \left( \frac{|\vec{p}_a|}{|\vec{p}_\gamma|} \right)^{2J+1} \nonumber 
\\
&\times& \left( \frac{ g_{aNN}^0 \beta + g_{aNN}^1}{(\mu_0-1/2)\beta + \mu_1 - \eta} \right)^2,
\end{eqnarray}
where $\beta$ and $\eta$ are nuclear structure factors.
The GEANT4 simulation of IsoDAR provides several transition lines including the energy and the flux, as shown in Fig. \ref{fig:GEANTplot}.
 The nuclear structure factors $\beta$ and $\eta$ are then computed in BIGSTICK~\cite{Johnson:2018hrx}, and we consult the~\href{https://www-nds.iaea.org/relnsd/vcharthtml/VChartHTML.html}{NDS}~(Nuclear Data Service) to find the transitions that match with the lines GEANT4 generates. See Supplemental Material at {\color{red}[URL by publisher]} for the selected lines. Since typical values of $\beta \simeq 1$ for the isotopes that source ALPs in the IsoDAR target, this makes the branching ratio in Eq.~\ref{eq:br} roughly proportional to $g_{aNN}^0 + g_{aNN}^1$, we take a single effective coupling $g_{ann} \equiv g^0_{aNN} + g^1_{aNN}$ to simplify our view of the parameter space. 

\noindent \textbf{ALP Detection Mechanisms.}  
We express the probability of detection in several steps:
\begin{enumerate}
  \item The probability that the ALP survives to the detector without decaying, $P_{S1}$.
  \item The probability that the ALP survives to the detector without scattering,  $P_{S2}$.
  \item The probability the ALP decays inside the detector to visible energy, $P_{decay}$.
  \item The probability the ALP scatters inside the volume of the detector, $P_{scatter}$.
\end{enumerate}

The probability of ALP decay can be calculated by integrating the probability density over the distance the ALP travels. In the case of the probabilities for which the ALP is producing photons in the detector, the integrated distances are the boundaries of the detector, or $l$. For survival probabilities, we will call this distance $d$. 
Therefore, the probability of the ALP not decaying before the detector, or decaying inside the detector are:
\begin{equation}\label{p_surv_dec}
P_{S1} = e^{-d/(\tau v_a)} ,  
P_{decay} = 1- e^{-l/(\tau v_a)}
\end{equation}

where $\tau = \gamma/\Gamma$ is the lab frame lifetime of the ALP $v_a= p_a/E$. For couplings to photons, for example, the decay width of the ALP is related to its mass and $g_{a \gamma}$ by 
$\Gamma = (g_{a\gamma}^2 m_a^3)/(64\pi)$ and $\gamma$ is the Lorentz factor.

Similarly, to find $P_{S2}$ and $P_{scatt}$, we can alter equation \ref{p_surv_dec} by replacing $1/(\tau v_a)$ with $n\sigma(E)$. Here $n$ is the number density of the material the ALP is traveling through and $\sigma$ is the cross section of the coupling being investigated.

The total probability is found by integrating over the solid angle of the detector.This integration is uniform due to the isotropic processes of  production.

\begin{align}
P_{total} = \int_{ \frac{-5\pi}{6}}^{\frac{5\pi}{6}}\int_{ \frac{-5\pi}{6}}^{\frac{5\pi}{6}} & P_{S1} P_{S2}
\bigg[P_{scatter}+ P_{decay} \nonumber \\
& - (P_{scatter}* P_{decay})\bigg] d\theta d\phi
\label{p_total}
\end{align}

\begin{figure}[ht]
    \centering
    \includegraphics[width=0.45\textwidth]{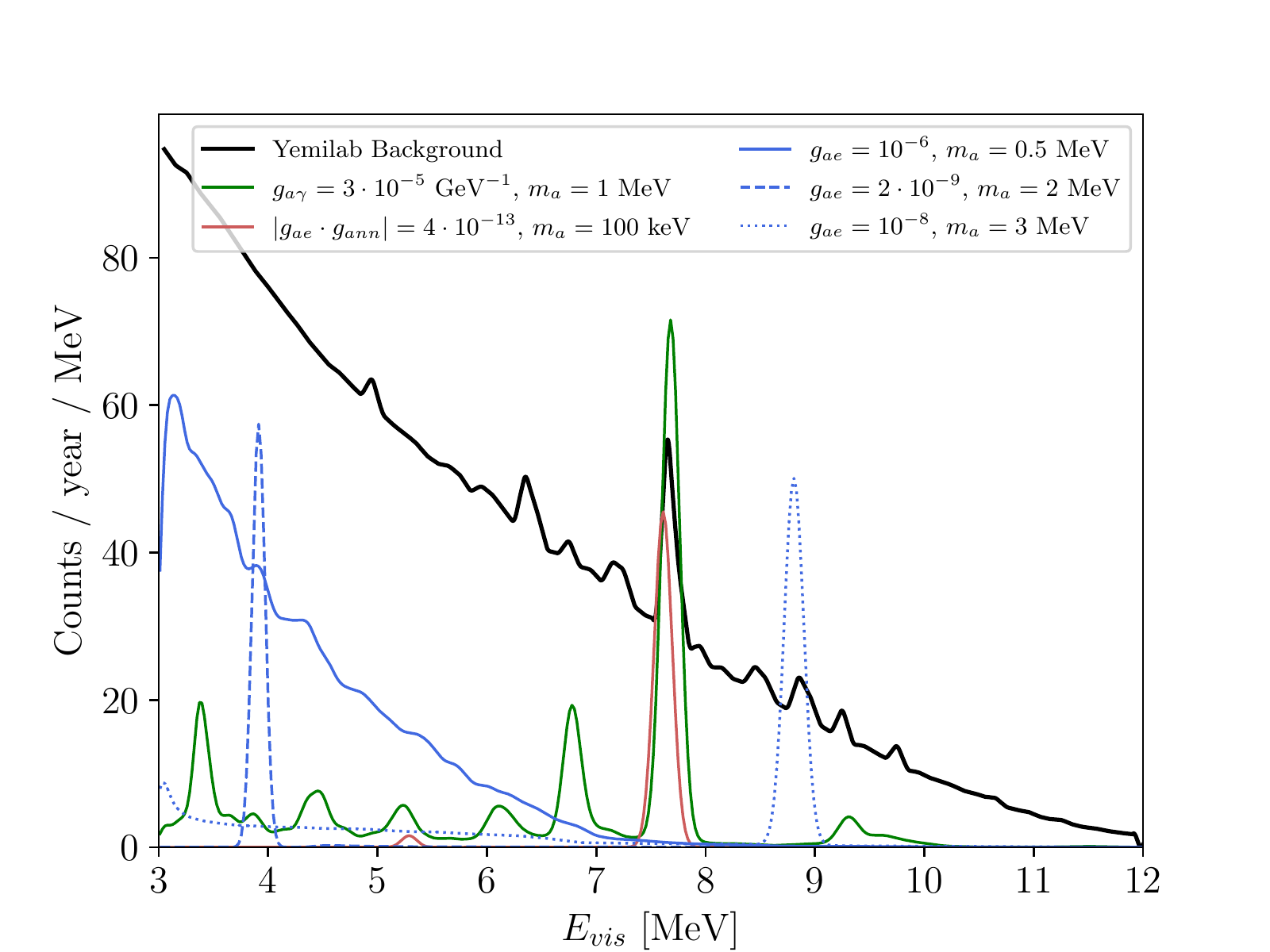}
    \caption{ALP production and detection in various coupling combinations. Nuclear transitions and emission lines in the photon flux showing up for $g_{a\gamma}$ and $g_{ann}$ dependent signals (red, green) while resonantly produced ALPs through $e^+ e^-$ to sharp peaks positioned at $\propto m_a^2$ (blue), provided that $m_a > 2 m_e$. The Yemilab background is shown in black for comparison. A detector energy resolution of 3\% is applied.}
    \label{fig:gae_spectra}
\end{figure}

 For ALPs with masses $m_a > 2 m_e$ the decay channel to $e^+ e^-$ becomes available, with width $\Gamma(a \to e^+e^-) = \frac{g_{ae}^2 m_a}{8\pi}(1 - \frac{4m_e^2}{m_a^2})^{1/2}$. Detection rates from $a \to e^+ e^-$ decays inside the detector fiducial volume can be calculated using the same probabilities from the previous section. Alternatively, ALPs with couplings to electrons can be detected via inverse-Compton scattering ($a e^- \to \gamma e^-$). The resulting detected energy spectra from electron coupling production and detection channels is shown in Fig.~\ref{fig:gae_spectra}.\\

The total number of ALP events requires a convolution of this probability density with the associated spectrum. We take $5\pi/6$ as the angle that the LSC detector covers from the IsoDAR target based on the geometry in ref.~\cite{alonso2022isodar}:
\begin{equation}
N_{events} = \int_{3~\mathrm{MeV}}^{\infty} S(E) P_{total}(E) dE
\label{eq:evt_rate}
\end{equation}
where $S(E)$ is the photon or electron spectrum weighted by the probability of ALP production in the target. A lower limit of 3 MeV was taken for the LSC detector as below this energy threshold additional backgrounds become insurmountable due to radiogenic-induced gammas. The signals in this letter are identical to electron-like events seen in the detector, and have a 32\% detection efficiency. In the LSC photon and electron-like signals are indistinguishable, and no angular reconstruction has been used in this case. These are explored more in ~\cite{Alonso:2021kyu}.

\noindent \textbf{Background Analysis.}
The IsoDAR backgrounds have been well studied. The LSC cannot distinguish electron-antineutrino-like events from ALP-like events, we thus use the background analysis from refs.~\cite{alonso2022isodar, alonso2022neutrino, Alonso_2022}, included in   Fig.~\ref{fig:gae_spectra}, while adding the neutrino events as a background to the ALP signal. These backgrounds include solar neutrinos, cosmogenic isotopes, radiogenics within the detector and from rock surrounding the detector.

Artificial noise was added in a Poisson distribution to each 0.1 MeV bin of the background spectrum to simulate random fluctuations in the data. 10,000 pseudo-experiments were thrown with and without ALPs injected into the data and the signal events were counted. A $\Delta\chi^2$ test was then used to compare signal with and without ALPs for each set of ALP parameters. We then compared the signal at this point in parameter space to background to see if it could be differentiated from background at 90\% confidence level. This process was repeated over the relevant ALP parameter space in order to draw the projected sensitivity.\\


\begin{figure*}[ht!]
\centering
\subfloat[]{
\includegraphics[width=0.45\textwidth]{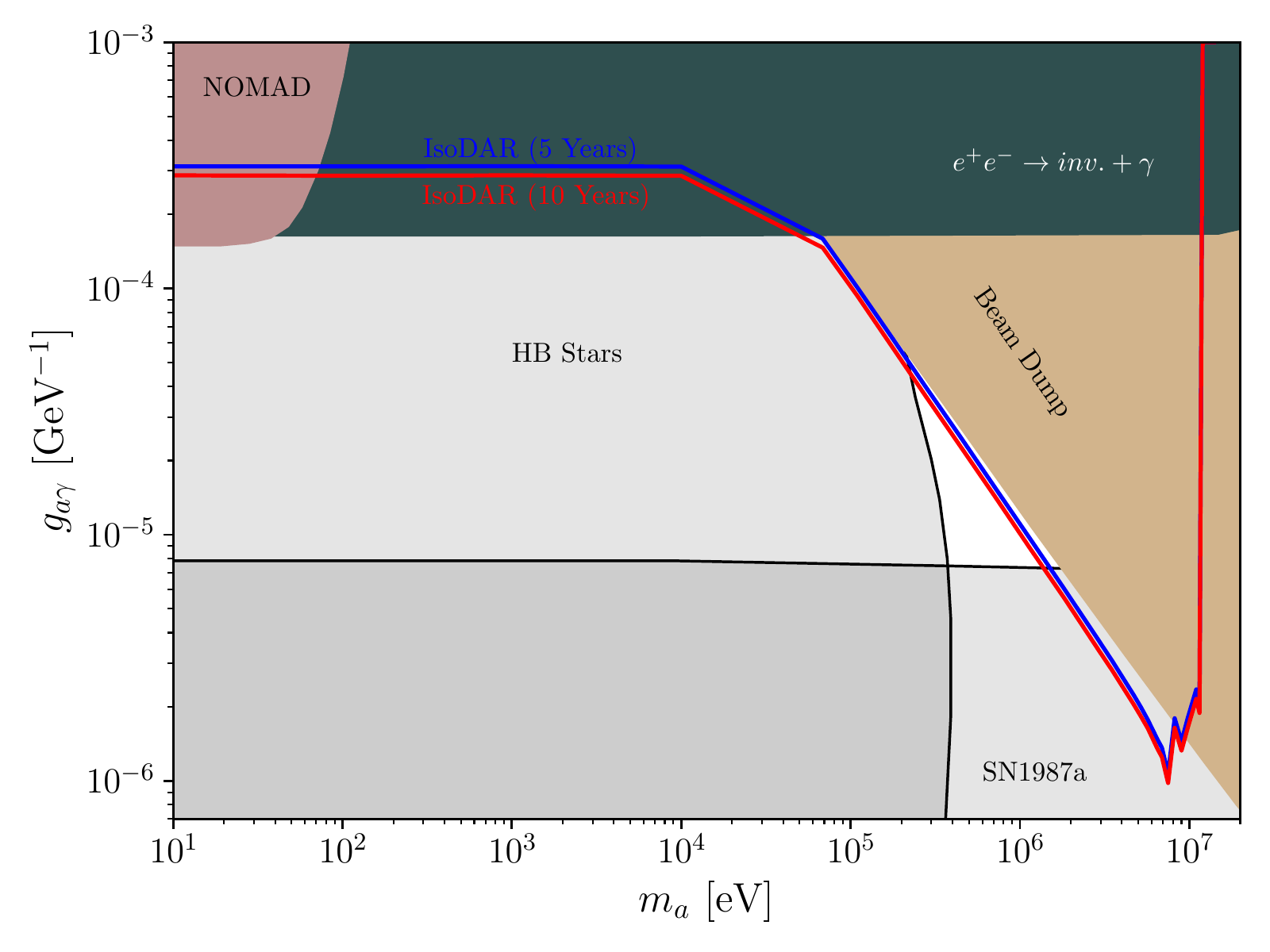}
} 
\subfloat[]{
\includegraphics[width=0.45\textwidth]{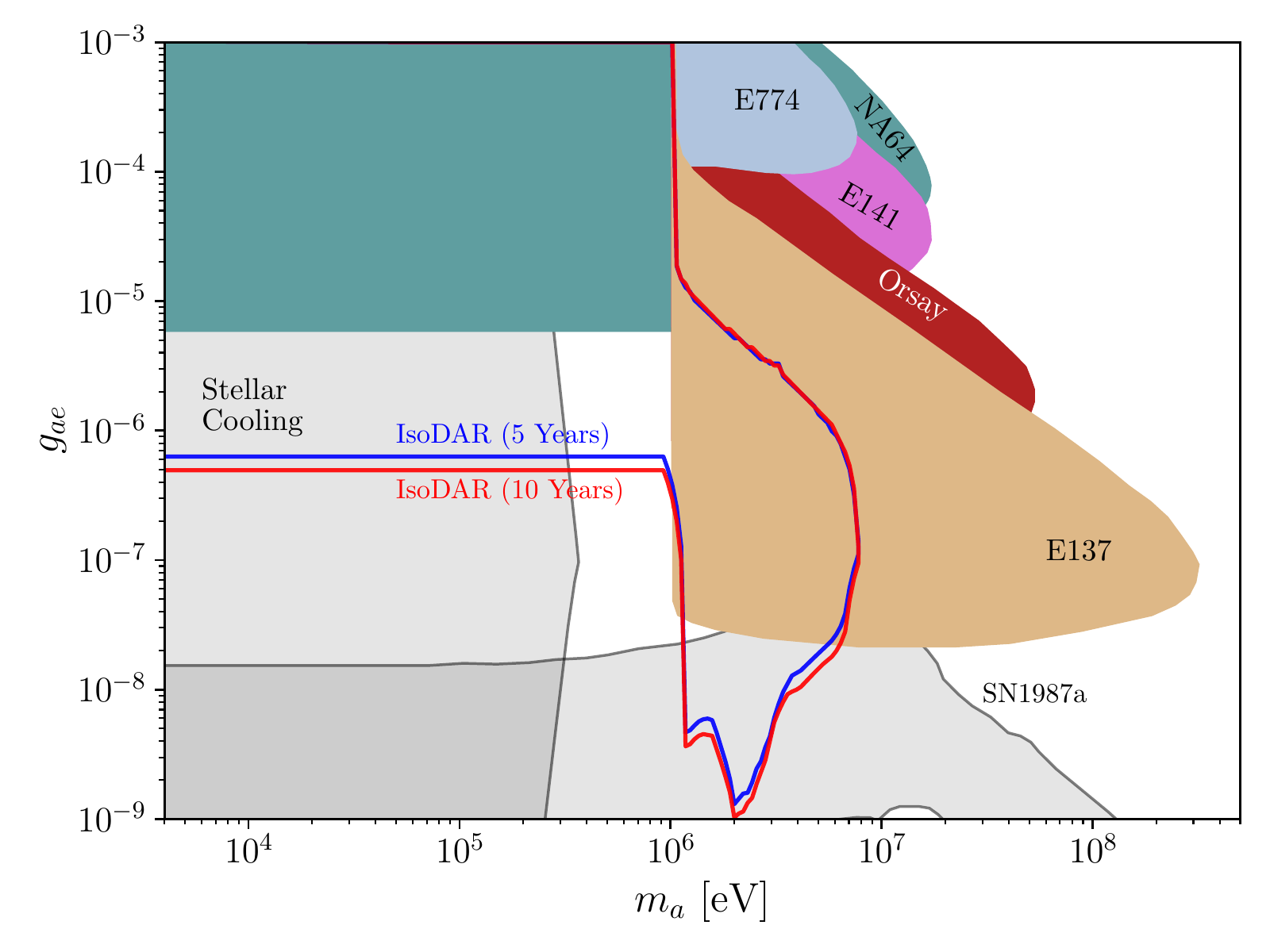}
}\\
\subfloat[]{
\includegraphics[width=0.45\textwidth]{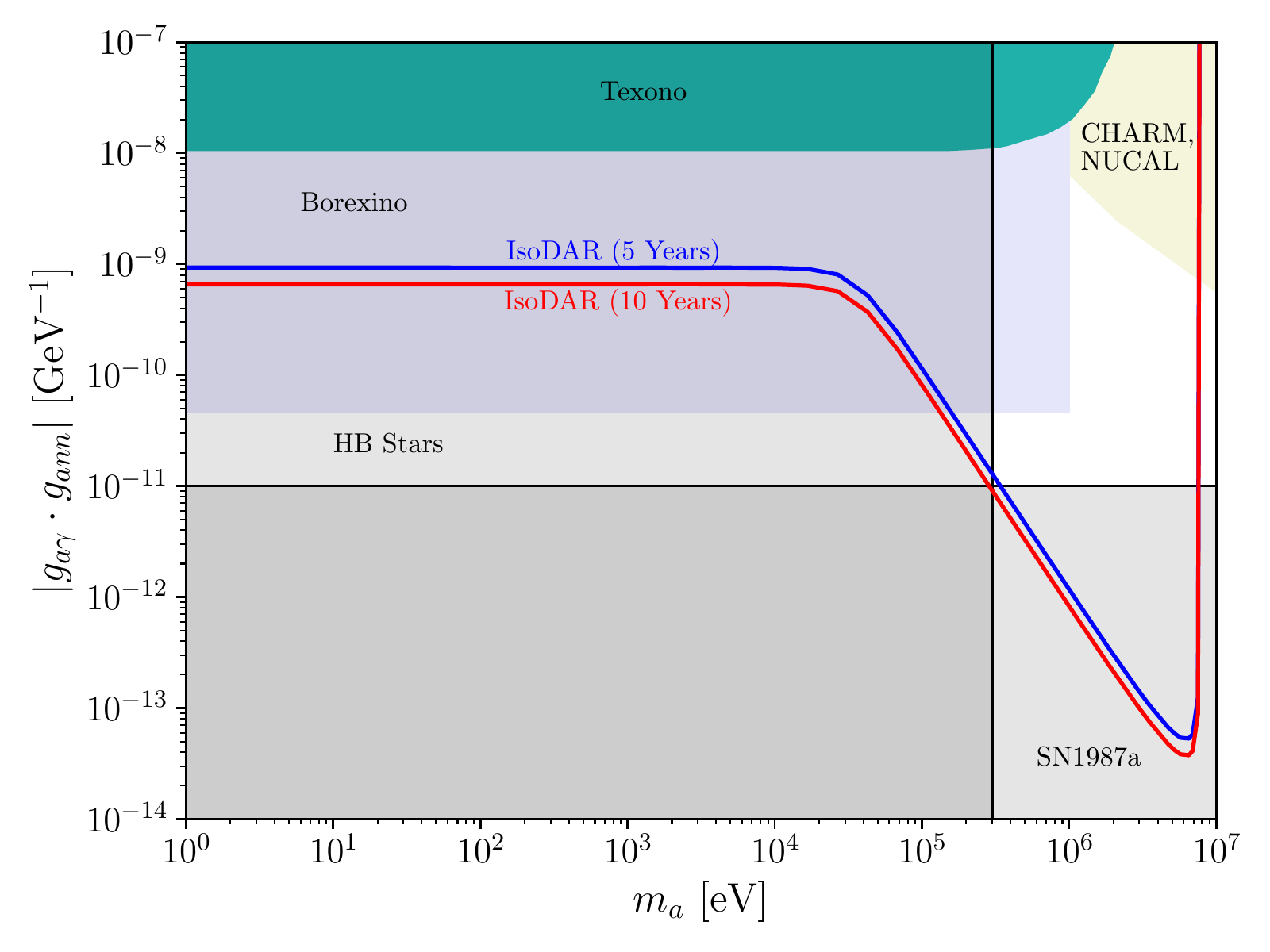}
}
\subfloat[]{
\includegraphics[width=0.45\textwidth]{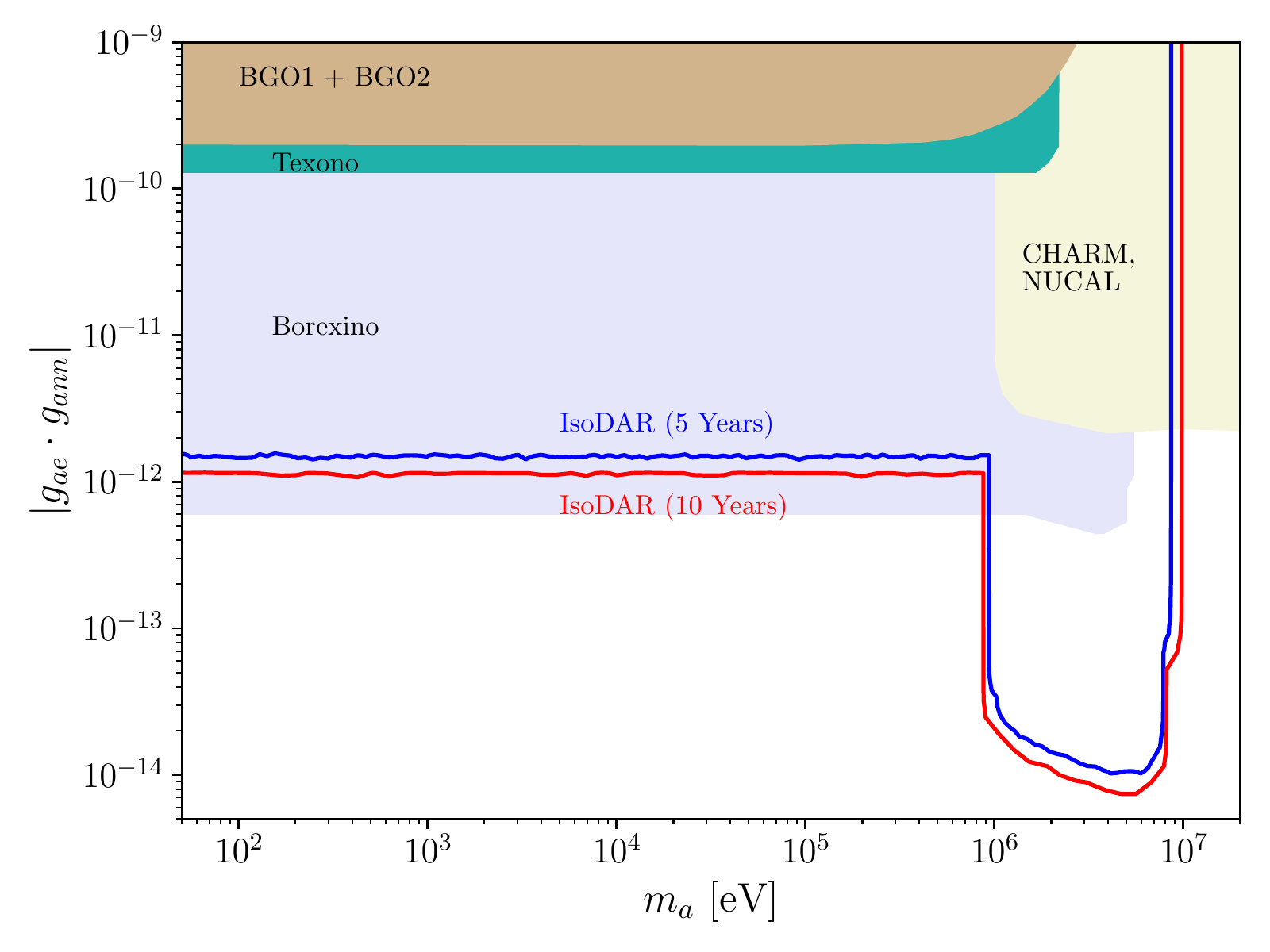}
}
\caption{
Sensitivity contours at 90\% CL, for 5 and 10 year exposures, using (a) couplings to photons, (b) couplings to electrons, (c) couplings to nucleons and photons, and (d) couplings to nucleons and electrons. In (c) and (d), ALPs are produced via nuclear transitions and propagate to the detector to subsequently scatter or decay via electron coupling (inverse Compton, $a\to e^+ e^-$ decay) or photon coupling (inverse Primakoff, $a \to \gamma\gamma$ decay) channels.}
\label{fig:4_exclusion}
\end{figure*}

\noindent \textbf{Results and Conclusions.}
Fig.~\ref{fig:4_exclusion} shows IsoDAR's projected sensitivity to ALP parameter space over nucleon, photon, and electron couplings using an estimated $7.88\cdot10^{24}$ protons on target over 5 years. We have presented the existing limits in model-independent way, but if the reader wishes to compare this parameter space to traditional QCD axion models (such as the Dine-Fischler-Srednicki-Zhitnitsky (DFSZ) type models~\cite{Zhitnitsky:1980tq,DINE1981199,Dine:1981rt,Dine:1982ah} or Kim-Shifman-Vainshtein-Zakharov (KSVZ)~\cite{PhysRevLett.43.103, SHIFMAN1980493} models), one can refer to refs.~\cite{Bhattarai:2022mue, CCM:2021lhc, AristizabalSierra:2020rom} where the relevant QCD model space is highlighted (although the effect of a multi-parameter analysis of a full QCD axion model on the single-parameter constraints has not been considered).


We first present the case where the production and detection mechanisms are solely via the photon coupling. The sensitivity is shown in Fig.~\ref{fig:4_exclusion}(a). We find a slight increase in the sensitivity over generic ALP parameter space, extending the existing beam dump constraints down in the coupling until the SN1987a and HB stars constraints~\cite{Caputo:2022mah,CARENZA2020135709}.\footnote{See also ref.~\cite{Caputo:2021rux} for a discussion on potential constraints from supernovae explosion energy which may apply in this region of parameter space.} 


The IsoDAR experiment tests astrophysical limits, which are known to be model-dependent (see, e.g., ref.~\cite{Bar:2019ifz}), and are often being revised with new data and theoretical guidance. They can also be lifted in several specific models~\cite{Jaeckel:2006xm, Khoury:2003aq, Brax:2007ak, Masso:2005ym, Masso:2006gc, Dupays:2006dp, Mohapatra:2006pv}, and in this sense IsoDAR probes valuable parameter space in a model-independent way.

The projected sensitivity to ALPs coupling to electrons is shown in Fig.~\ref{fig:4_exclusion}(b), where we have considered ALP production in the IsoDAR target via Compton scattering, associated and resonant production, and ALP-bremsstrahlung. We project sensitivity over new values of $g_{ae}$ beyond the missing-energy constraint from NA64~\cite{NA64:2021ked,Andreev:2021fzd,Gninenko:2017yus}, and even extend the existing decay limits beyond the beam dump bounds~\cite{Bechis:1979kp,PhysRevD.38.3375,Andreas:2010ms,Riordan:1987aw,Bross:1989mp} and test the space only excluded by the SN1987a neutrino measurement~\cite{Lucente:2021hbp}.

Next we show a sensitivity projection in the combined parameter space of nucleon, photon, and electron couplings. For simplicity we assume ALPs are produced primarily in nuclear transitions and then propagate to the detector to either decay or scatter through either $g_{ae}$ or $g_{a\gamma}$ mediated channels. We perform the parameter space scan in a similar way to ref.~\cite{AristizabalSierra:2020rom} by iterating over combinations of $(g_{ann}, g_{ae})$ or $(g_{ann}, g_{a\gamma})$ and mapping CL contours in the product space $|g_{ann}\cdot g_{ae}|$ or $|g_{ann}\cdot g_{a\gamma}|$. Note there is a sharp cutoff at $\sim$10 MeV from where the photon spectrum terminates as seen in Fig.~\ref{fig:GEANTplot}. The flat regions at lower masses describe the regime in which detection is scattering dominated, while at higher masses the detection becomes decay dominated. The "wiggles" at higher masses in Fig.~\ref{fig:4_exclusion}(a) are due to the ALP masses exceeding the photon peaks seen in Fig.~\ref{fig:GEANTplot}, causing sudden drops in production. The sensitivity for ALPs produced via nuclear transitions and detected by inverse Primakoff scattering and decays to $\gamma\gamma$ ($g_{ae} \to 0$, $g_{a\gamma} \neq 0$, $g_{ann} \neq 0$) can be seen in Fig.~\ref{fig:4_exclusion}(c), which projects a new exclusion over almost a full order of magnitude in the coupling product $|g_{a\gamma} \cdot g_{ann}|$, while also testing astrophysical constraints. The sensitivity reach using the same production modes but with detection through inverse Compton scattering and decays to $e^+ e^-$ ($g_{a\gamma} \to 0$, $g_{ae} \neq 0$, $g_{ann} \neq 0$) is shown in Fig.~\ref{fig:4_exclusion}(d), and also shows good reach over new parameter space for $m_a > 2m_e$ where the $e^+ e^-$ decays become kinematically accessible. Existing bounds from Borexino~\cite{Borexino:2008wiu, Borexino:2012guz}, Texono~\cite{TEXONO:2006spf}, and BGO bolometers~\cite{Derbin:2013zba, Derbin:2014xzr} are shown along with the astrophysical bounds from HB stars and SN1987a in both scenarios in Figs.~\ref{fig:4_exclusion}(c) and~\ref{fig:4_exclusion}(d). In addition, the null observations of CHARM~\cite{BERGSMA1985458} and NUCAL~\cite{Blumlein:1990ay} may be used to set constraints by considering ALP production from proton bremsstrahlung in their beam targets through the nuclear couplings $g_{aNN}^0$, $g_{aNN}^1$ and decays to $\gamma\gamma$ and $e^+e^-$. We have computed the exclusions from CHARM and NUCAL independently in this work and show them in Figs.~\ref{fig:4_exclusion}(c,d).


In conclusion, we have presented a strategy to study ALPs utilizing nuclear transitions in a fixed target neutrino facility. These monoenergetic peaks provide specific signal above background leading to probing of parameter space which would otherwise require higher flux. These peaks provide new ways to probe axion-nucleon couplings. We used the IsoDAR neutrino experiment as an example. Using GEANT4 we showed the IsoDAR target provides monoenergetic peaks from nuclear excitations, as well as a high flux of electrons that can be used to study electron-ALP couplings. Using this technique we were able to demonstrate that this experiment will be able to duplicate lab based and astrophysical constraints, as well as probe entirely new parameter spaces that span several models.  This study has focused on use of the LSC detector, as this is planned for other IsoDAR physics goals; however, in the future, one could consider pairing the source with a detector optimized for ALP searches, such as a detector who has thresholds below 3 MeV, or  substitute the IsoDAR neutrino target with one that has been optimized to produce ALPs through nuclear excitations, rather than beta-decay. The backgrounds of an ALP experiment would be reduced with the neutrino flux, further improving the sensitivities presented here.
\\

\section*{Acknowledgement}
BD and AT acknowledge support from the U.S. Department of Energy (DOE) Grant DE-SC0010813. The work of DK is supported by DOE under Grant No. DE-FG02-13ER41976/DE-SC0009913/DE-SC0010813. We acknowledge that portions of this research were conducted with the advanced computing resources provided by Texas A\&M High Performance Research Computing. LW, AB. JC, MS, and JS had work that was supported by NSF grants PHY-1912764 and
PHY-1626069, and the Heising-Simons Foundation.

\appendix
\section{QCD Axion Benchmark Models and their Parameter Space}
The correlations between the QCD axion mass and its effective couplings are given below, taken from ref.~\cite{DiLuzio:2020wdo}. We simply reiterate those correlations here for convenience of the reader. The relation between the Peccei-Quinn breaking scale $f_a$ and the axion mass is
\begin{equation}
    f_a =  \bigg(\frac{5.691\times 10^{6} \textrm{eV}}{m_a}\bigg) \textrm{GeV}
\end{equation}
To find the correlations between the axion mass and its effective couplings to photons in the Kim-Shifman-Vainshtein-Zakharov (KSVZ) benchmark model is then given by Eq.~\ref{eq:ksvz_gagamma};
\begin{align}
\label{eq:ksvz_gagamma}
    g_{a\gamma} &= \frac{m_a}{\textrm{GeV}} \bigg(0.203 \frac{E}{N} - 0.39\bigg)
\end{align}
We then consider a range of model parameter space by considering anomaly number ratios of $E/N = 44/3$ to $E/N = 2$. This defines a band in $(m_a, g_{a\gamma}$ parameter space in which the QCD axion's couplings and mass may reside.

For the Dine-Fischler-Srednicki-Zhitnitsky (DFSZ) benchmark model, for which couplings to electrons would be dominant relative to the photon couplings, we take
\begin{equation}
\label{eq:dfsz_gae}
    g_{ae} = \dfrac{m_e C_{ae}(m_a, \tan\beta)}{f_a}
\end{equation}
where the coefficient $C_{ae}$ is dependent on the rotation angle $\beta$ for the vacuum expectation values of the extended Higgs sector in DFSZI and DFSZII models;
\begin{multline}
\textrm{DFSZ(I):\, \, }C_{ae} = -\frac{1}{3} \sin^2\beta + \frac{3\alpha^2}{4 \pi^2} \bigg[\frac{E}{N}  \log(f_a/m_a) \\ - 1.92 \log(1/m_e)\bigg],  \, \, \, \,   \frac{E}{N} = 8/3 \end{multline}

\begin{multline}
\textrm{DFSZ(II):\, \,  } C_{ae} = \frac{1}{3} \sin^2\beta + \frac{3\alpha^2}{4 \pi^2} \bigg[\frac{E}{N}  \log(f_a/m_a) \\ - 1.92 \log(1/m_e)\bigg], \, \, \, \,     \frac{E}{N} = 2/3
\end{multline}

Here we take $\tan\beta$ values between 0.25 and 120, which equates to $\sin\beta = 0.242536$ and  $\sin\beta = 0.999965$, respectively~\cite{Giannotti:2017hny}.

\section{Nuclear Excited States in the IsoDAR Target and ALP Event Rates}

The nuclear excited states induced through proton collision and particle shower cascades in the IsoDAR target are listed for the M1 and M2 transitions relevant for pseudoscalar ALP production in Table~\ref{tab:translines}. The best matches of nuclear transitions between GEANT4 and NDS are presented for select isotopes. Recall the ALP production branching ratio given by ref.~\cite{Avignone:1988bv} for a transition is
\begin{equation}
\begin{split}
\left(\frac{\Gamma_a}{\Gamma_\gamma}\right)_{\text{MJ}} &= \frac{1}{\pi \alpha} \frac{1}{1+\delta^2} \frac{J}{J+1} \\
&\quad\times \left(\frac{|\vec{p}_a|}{|\vec{p}_\gamma|}\right)^{2J+1} \left(\frac{g_{aNN}^0 \beta + g_{aNN}^1}{(\mu_0-1/2)\beta + \mu_1 - \eta}\right)^2,
\end{split}
\end{equation}

Nuclear structure factors $\beta$ and $\eta$ are computed by BIGSTICK~\cite{Johnson:2018hrx} if possible, otherwise $\beta$ and $\eta$ are set to the default values (1 and 0.5, respectively).

\begin{table}[h]
\centering
\begin{tabular}{|l|l|l|l|l|}
    \hline
    \hline
    Nucleus & Energy in MeV & Type & $\beta$ & $\eta$ \\
    \hline
    Fe57 & 7.606 & M1 & 0.7071 & -0.3111 \\
    \hline
    Li8 & 1.009 & M1 & 1 & -0.0260 \\
    \hline 
    Li8 & 2.053 & M1 & 1 & -0.1034 \\
    \hline
    O15 & 5.281 & M2 & 1 & 0.5 \\
    \hline
    \hline
\end{tabular}
\caption{M1 and M2 nuclear transitions, their energies, and nuclear parameter inputs to the ALP production branching ratio.}
\label{tab:translines}
\end{table}

\bibliography{apssamp}

\end{document}